\documentstyle[12pt]{article}
\newfont{\Mb}{msbm10}

\begin{document}
\setcounter{equation}{0}
\setcounter{figure}{0}
\setcounter{table}{0}

\title{A Method to Tackle First Order Ordinary 
Differential Equations with Liouvillian Functions in the Solution - II}
\author{
L.G.S. Duarte\thanks{
     Universidade do Estado do Rio de Janeiro,
     Instituto de F\'{\i}sica, Departamento de  F\'{\i}sica Te\'orica,
     R. S\~ao Francisco Xavier, 524, Maracan\~a, CEP 20550--013,
     Rio de Janeiro, RJ, Brazil. E-mail: lduarte@dft.if.uerj.br},
S.E.S. Duarte\thanks{
     idem. E-mail: sduarte@dft.if.uerj.br},      
and L.A.C.P. da Mota\thanks{
     idem. E-mail: damota@dft.if.uerj.br}
}

\maketitle

\abstract{We present a semi-decision procedure to tackle first order
differential equations, with Liouvillian functions in the solution
(LFOODEs). As in the case of the Prelle-Singer procedure, this method
is based on the knowledge of the integrating factor structure.}

\newpage
\section{Introduction}

The problem of solving ordinary differential equations (ODEs) has led,
over the years, to a wide range of different methods for their solution.
Along with the many techniques for calculating tricky integrals, these
often occupy a large part of the mathematics syllabuses of university
courses in applied mathematics round the world. 

For a first order
differential equation (FOODE), finding the
solution can be equated to determining an integrating factor.
A remarkable method for finding such factors was developed, in 1983, by
Prelle and Singer \cite{PS}. Their method is based on the knowledge of
the general structure of the integrating factor for FOODEs of the type
$dy/dx=M(x,y)/N(x,y)$, with $M$ and $N$ polynomials in their arguments,
which present a solution that can be written in terms of elementary
functions (EFOODEs)\footnote{For a formal definition of elementary function, see
\cite{davenport}.}. Their approach is very attractive due to the fact
that it is non-classificatory and of a semi-decision nature. Therefore,
it has motivated many extensions of the original idea
\cite{Shtokhamer,collins,ManMac,PS2}.

Using the results presented in \cite{singer}, we presented (see \cite{firsTHEOps1}) a method which is an extension to the Prelle-Singer (PS) procedure allowing for the solution of some LFOODEs\footnote{Liouvillian functions are an extension of elementary functions, see \cite{davenport}.}
($dy/dx=M(x,y)/N(x,y)$, with $M$ and $N$ polynomials in their arguments). The
method is also based on the general structure of the integrating factor,
that was concluded to be of the form: $R = e^{r_0(x,y)} \prod_{i=1}^{n}
v_i(x,y)^{c_i}$, where $r_0$ is a rational function of $(x,y)$, the
$v_i$'s are irreducible polynomials in $(x,y)$ and the $c_i$'s are
constants. 

The method presented on \cite{firsTHEOps1} used a conjecture about the nature of the $v_i$'s (proved on \cite{secondTHEOps1}) and was restricted to a class of
LFOODEs, namely the ones to which $r_0(x,y)$ is either $f(x)$ or $g(y)$ or
$f(x)+g(y)$, where $f, g$ are rational functions. Here, we further
detail the structure of the integrating factor thus allowing us to
remove the above mentioned restriction about $r_0$, maintaining the semi-decision nature of the approach.

The paper is organized as follows: in section~\ref{Structure}, we
analyze the structure of the integrating factor for LFOODEs; in the
following section, we show how to apply that knowledge to construct a
semi-decision method to tackle LFOODEs and show some examples of
the application of the method. Finally, we present our conclusions.

\section{The Structure of the Integrating Factor}
\label{Structure}

Based on earlier results \cite{}, we will, in this section, improve our
knowledge of the structure of the integrating factor thus setting the
stage for the presentation of a semi-decision method to deal with LFOODEs. 

Let us then summarize these initial results:

\subsection{Previous Results}
\label{firstresults}

A seminal result on dealing with LFOODEs was obtained by Prelle and Singer in 1983 \cite{PS}. They have demonstrated that, for a LFOODE
\begin{equation}
\label{FOODE}
{\frac{dy}{dx}} = {\frac{M(x,y)}{N(x,y)}}
\end{equation}
where $M$ and $N$ are polynomials in $(x,y)$ with coefficients in the complex 
field $\it C$, if its solution can be written in terms of elementary functions, then there exists an integrating factor of the form $R = \prod_i f^{n_i}_i$ where $f_i$ are irreducible polynomials and $n_i$ are non-zero rational numbers. Appplying this to FOODEs of the type (\ref{FOODE}), we have:
\begin{equation}
\label{eq_PS}
{\frac{D[R]}{R}}  = \sum_i {\frac{n_iD[f_i]}{f_i}}=- \left( \partial_x N + \partial_y M\right),
\end{equation}
where $D \equiv N \partial_x + M \partial_y$.

From~(\ref{eq_PS}), plus the fact that $M$ and $N$ are polynomials, they
concluded that ${D[R]}/{R}$ is a polynomial and that $f_i |
D[f_i]$\footnote{In other words, $f_i$ is a factor of $D[f_i]$.} \cite{PS}. We now have a criterion
for choosing the possible $f_i$ (build all the possible divisors of
$D[f_i]$ up to a certain degree) and, if we manage to
solve~(\ref{eq_PS}), thereby finding $n_i$, we know the integrating
factor for the FOODE and the problem is reduced to a quadrature. 

In \cite{singer,firsTHEOps1,secondTHEOps1}, next steps were taken: it
was shown that, for a LFOODE of type (\ref{FOODE}), the integrating
factor is of the form: 

\begin{equation} 
\label{ourR}
R = e^{r_0(x,y)}  \prod_{i=1}^{n} v_i(x,y)^{c_i}.
\end{equation} 
where $r_0$ is a rational function of $(x,y)$, the $v_i$'s are
irreducible polynomials in $(x,y)$ and the $c_i$'s are constants. 

From this, we could conclude (see \cite{secondTHEOps1}) that $D[r_0]$ is a
polynomial and that the $v_i$'s are eigenpolynomials of the $D$
operator.

\subsection{A Theorem Concerning the Structure of $r_0$}
\label{theorem}

In this section, we are going to demonstrate a result about the structure of $r_0$ that will allow us to generalize the method presented in \cite{firsTHEOps1}. In order to do so, we are going to use some earlier results\footnote{These results can be found in 
\cite{firsTHEOps1,secondTHEOps1}.}. 

For a LFOODE of the form $dy/dx=M(x,y)/N(x,y)$, where $M$
and $N$ are polynomials in $(x,y)$, the integrating factor $R$ is given by $R = e^{r_0(x,y)}  \prod_{i=1}^{n} v_i(x,y)^{c_i}$, where $r_0$ is rational function of $(x,y)$, $v_i$ are irreducible eigenpolynomials of the $D$ operator (where $D \equiv N \partial_x + M \partial_y$)$, c_i$ are constants and $D[r_0]$ is a polynomial in $(x,y)$.

\bigskip
\bigskip

{\bf Theorem 1:} {\it Let the exponent $r_0$ be expressed as $P(x,y)/Q(x,y)$, where $P$
and $Q$ are polynomials in $(x,y)$, with no common factor. Then we have that $Q|D[Q]$ $($i.e., $D[Q]/Q$ is a polynomial 
in $(x,y))$.}

\bigskip

{\bf Proof:}
Since $D[r_0]$ is polynomial (see \cite{secondTHEOps1}), we can write it as:

\begin{equation}
\label{EQ_DRzero}
D[r_0] = D\left[\frac{P}{Q}\right] = \frac{Q\,D[P] - P\,D[Q]}{Q^2} = \Pi
\end{equation}
where $\Pi$ is polynomial in $(x,y)$. Multiplying (\ref{EQ_DRzero}) by $Q$, one obtains:

\begin{equation}
\label{EQ_DRzeroQ}
D[P] - \frac{P\,D[Q]}{Q} = \Pi Q.
\end{equation}

Since $D$ is a linear differential operator, with polynomial coefficients, and $P$ is polynomial, $D[P]$ is also polynomial. Therefore, since $\Pi\,Q$ is also polynomial, we may conclude that $\frac{P\,D[Q]}{Q}$ is polynomial either. Since, by hypothesis, $P$ and $Q$ have no common factor, we can infer that $Q|D[Q]$ (i.e., $D[Q]/Q$ is a polynomial), as we wantted to demonstrate.

\bigskip
\bigskip
This result, in turn, leads to the following corollary:

\bigskip

{\bf Corollary 1:} {\it We can write $Q$ as $\prod_{i=1} q_i(x,y)^{m_i}$, where the $q_i$'s are irreducible independent eigenpolynomials of the $D$ operator and the $m_i$'s are positive integers.}

\bigskip
\bigskip

{\bf Proof:} $Q$ is a polynomial so it can be written as $\prod_{i=1} q_i(x,y)^{m_i}$, where the $q_i$'s are independent irreducible polynomials and the $m_i$'s are positive integers. So, we can write:

\begin{equation}
\label{DQoverQ}
{\frac{D[Q]}{Q}} =  \sum_{i} m_i{\frac{D[q_i]}{q_i}}, 
\end{equation}
implying that $q_i|D[q_i]$ (i.e., $D[q_i]/q_i$ is a polynomial), as we wantted to show.

\section{A Semi-Decision Method to Tackle LFOODEs}
\label{method}

In this section, we will show how to construct, in a semi-decision way (as in the case of the
PS-method for EFOODEs), a method to deal with LFOODEs of the form $dy/dx=M(x,y)/N(x,y)$, where $M$ and $N$ are polynomials in $(x,y)$.

\subsection{Introduction}
\label{introM}

We now known that the integrating factor for these LFOODEs is of the form,

\begin{equation}
\label{ourR2}
R = e^{P(x,y)/Q(x,y)}\,S(x,y),
\end{equation} 
where we have that:
\begin{itemize}
\item $P$ and $Q$ are polynomials in $(x,y)$.
\item $Q$ can be written as $\prod_{i} \left(v_{\!\hbox{\footnotesize {\it q}}\,i}\right)^{\,m_i}$, where the $v_{\!\hbox{\footnotesize {\it q}}}$'s are independent irreducible eigenpolynomials (of the $D$ operator) and the $m$'s are positive integers.
\item $S$ can be written as $\prod_{j} \left(v_{\!\hbox{\footnotesize {\it s}}\,j}\right)^{\,c_j}$, where the $v_{\!\hbox{\footnotesize {\it s}}}$'s are independent irreducible eigenpolynomials (of the $D$ operator) and the $c$'s are positive integers.
\end{itemize}

In what follows, let us convert this knowledge into a form that will allow us to build a solving method.

Using (\ref{ourR2}) into $D[R]/R=-(\partial_x N + \partial_y M )$, we get:
\begin{equation}
\label{eqadendum1}
D\left[\frac{P}{Q}\right] + \frac{D[S]}{S} =  - \left( \partial_y  M+ \partial_x N \right),  
\end{equation}
leading to
\begin{equation}
\label{eqadendum2}
\frac{Q\,D[P]-P\,D[Q]}{Q^2} + \sum_j c_j{\frac{D[v_{\!\hbox{\footnotesize {\it s}}\,j}]}{v_{\!\hbox{\footnotesize {\it s}}\,j}}} =   - \left( \partial_y  M+ \partial_x N \right).
\end{equation}
Remembering that the $v_{\!\hbox{\footnotesize {\it s}}}$'s are eigenpolynomials of the $D$ operator, we can write $D[v_{\!\hbox{\footnotesize {\it s}}\,j}]=\lambda_{\!\hbox{\footnotesize {\it s}}\,j}\,v_{\!\hbox{\footnotesize {\it s}}\,j}$, where the $\lambda_{\!\hbox{\footnotesize {\it s}}}$'s are the polynomial `eigenvalues' in $(x,y)$ associated with the $v_{\!\hbox{\footnotesize {\it s}}}$'s. Thus, we can write (\ref{eqadendum2}) as:
\begin{equation}
\label{eqadendum3}
\frac{1}{Q}\,\left(D[P] - P\,\frac{D[Q]}{Q}\right) + \sum_j c_j\,\lambda_{\!\hbox{\footnotesize {\it s}}\,j} =   - \left( \partial_y  M+ \partial_x N \right).
\end{equation}
If we then multiply both sides of (\ref{eqadendum3}) by $Q$, we obtain:
\begin{equation}
\label{eqadendum4}
D[P] - P\,\frac{D[Q]}{Q} + Q\,\sum_j c_j\,\lambda_{\!\hbox{\footnotesize {\it s}}\,j} =   - Q\,\left( \partial_y  M+ \partial_x N \right).
\end{equation}
Since $Q = \prod_{i} \left(v_{\!\hbox{\footnotesize {\it q}}\,i}\right)^{\,m_i}$, where the $v_{\!\hbox{\footnotesize {\it q}}}$'s are independent irreducible eigenpolynomials of the $D$ operator, we can write:
\begin{equation}
\frac{D[Q]}{Q} = \sum_i\,m_i\,\frac{D[v_{\!\hbox{\footnotesize {\it q}}\,i}]}{v_{\!\hbox{\footnotesize {\it q}}\,i}} = \sum_i\,m_i\,\lambda_{\!\hbox{\footnotesize {\it q}}\,i},
\end{equation}
where the $\lambda_{\!\hbox{\footnotesize {\it q}}}$'s are the polynomial `eigenvalues' associated with the $v_{\!\hbox{\footnotesize {\it q}}}$'s. Using this into (\ref{eqadendum4}) and re-arranging we get:
\begin{equation}
\label{eqadendum5}
D[P] - P\,\sum_i\,m_i\,\lambda_{\!\hbox{\footnotesize {\it q}}\,i} = - \prod_{i} \left(v_{\!\hbox{\footnotesize {\it q}}\,i}\right)^{\,m_i}\,\left(\sum_j c_j\,\lambda_{\!\hbox{\footnotesize {\it s}}\,j} + \partial_y  M+ \partial_x N \right).
\end{equation}

This equation will prove to be very important to our method. So, let us now do some analysing of its structure: We don't know the degree of the eigenpolynomials $v_{\!\hbox{\footnotesize {\it s}}}$ that build up the function $S$. Neither do we know the degree of the eigenpolynomials $v_{\!\hbox{\footnotesize {\it q}}}$ that appear on $Q$. However, once these degrees are set and choosing the values for the exponents $m_i$ (thus completing the determination of the degree of $Q$), we can determine the degree of the polynomial $P$. Let us see how it goes by analysing both sides of (\ref{eqadendum5}).  

First, note that $D[P] = M\,\partial_y P + N\,\partial_x P$. Since the degree of $\partial_x P$ is at most equal to the degree of $P$ minus $1$, and the same is valid for the degree of $\partial_y P$, we may conclude that (in what follows, we will denote the degree of any term by $\hbox{\tt d}_{\hbox{\scriptsize term}}$):
\begin{equation}
\label{degree1}
\hbox{\tt d}_{\hbox{\scriptsize {\it D}[{\it P}]}} \, \leq \, \max (\hbox{\tt d}_{\hbox{\scriptsize {\it M}}},\hbox{\tt d}_{\hbox{\scriptsize {\it N}}}) + \hbox{\tt d}_{\hbox{\scriptsize {\it P}}} -1.
\end{equation}
Looking now to the second term of the left hand size of (\ref{eqadendum5}) we may say that
\begin{equation}
\label{degree2}
\hbox{\tt d}_{{\hbox{\scriptsize {\it P}}} \sum m_i \lambda_{\!\hbox{\scriptsize {\it q}}\,i}} \, \leq \, \max (\hbox{\tt d}_{\lambda_{\!\hbox{\scriptsize {\it q}}}}) + \hbox{\tt d}_{\hbox{\scriptsize {\it P}}}.
\end{equation}
But we know that  $D[v_{\!\hbox{\footnotesize {\it q}}\,i}]=\lambda_{\!\hbox{\footnotesize {\it q}}\,i}\,v_{\!\hbox{\footnotesize {\it q}}\,i}$, leading to:
\begin{equation}
\label{degree3}
\max (\hbox{\tt d}_{\lambda_{\!\hbox{\scriptsize {\it q}}}}) + \hbox{\tt d}_{v_{\!\hbox{\scriptsize {\it q}}}} \, \leq \, \max (\hbox{\tt d}_{\hbox{\scriptsize {\it M}}},\hbox{\tt d}_{\hbox{\scriptsize {\it N}}}) + \hbox{\tt d}_{v_{\!\hbox{\scriptsize {\it q}}}} - 1
\end{equation}
and, consequently, to 
\begin{equation}
\label{degree4}
\max (\hbox{\tt d}_{\lambda_{\!\hbox{\scriptsize {\it q}}}}) \, \leq \, \max (\hbox{\tt d}_{\hbox{\scriptsize {\it M}}},\hbox{\tt d}_{\hbox{\scriptsize {\it N}}}) - 1.
\end{equation}
Using (\ref{degree4}) into (\ref{degree2}) we can write
\begin{equation}
\label{degree5}
\hbox{\tt d}_{{\hbox{\scriptsize {\it P}}} \sum m_i \lambda_{\!\hbox{\scriptsize {\it q}}\,i}} \, \leq \, \max (\hbox{\tt d}_{\hbox{\scriptsize {\it M}}},\hbox{\tt d}_{\hbox{\scriptsize {\it N}}}) + \hbox{\tt d}_{\hbox{\scriptsize {\it P}}} -1.
\end{equation}
Through (\ref{degree1}) and (\ref{degree5}) we see that the maximum degree of the left rand side of (\ref{eqadendum5}) is $\max (\hbox{\tt d}_{\hbox{\scriptsize {\it M}}},\hbox{\tt d}_{\hbox{\scriptsize {\it N}}}) + \hbox{\tt d}_{\hbox{\scriptsize {\it P}}} -1$. Now, looking at the right hand size of (\ref{eqadendum5}), we have that its degree is at most equal to $\hbox{\tt d}_{\hbox{\scriptsize {\it Q}}} + \max (\hbox{\tt d}_{\lambda_{\!\hbox{\scriptsize {\it q}}}},{\hbox{\tt d}_{\partial_y\hbox{\scriptsize {\it M}}}},{\hbox{\tt d}_{\partial_x\hbox{\scriptsize {\it N}}}})$. Taking (\ref{degree4}) into account and the fact that ${\hbox{\tt d}_{\partial_y\hbox{\scriptsize {\it M}}}}\,\leq \,\hbox{\tt d}_{\hbox{\scriptsize {\it M}}} - 1$ and ${\hbox{\tt d}_{\partial_x\hbox{\scriptsize {\it N}}}}\,\leq \,\hbox{\tt d}_{\hbox{\scriptsize {\it N}}} - 1$, the maximum degree of the right hand size of (\ref{eqadendum5}) is $\hbox{\tt d}_{\hbox{\scriptsize {\it Q}}} + \max (\hbox{\tt d}_{\hbox{\scriptsize {\it M}}},\hbox{\tt d}_{\hbox{\scriptsize {\it N}}}) - 1$. Since (\ref{eqadendum5}) has to be satisfied, we see that the degree of $Q$ bounds that of $P$.
%

\subsection{The Steps of the Method}
\label{steps}

Based on what has been explained on section (\ref{introM}), let us now stablish the actual procedure to be followed. In equation (\ref{eqadendum5}) (the cornerstone of our method), 
we see terms involving the eigenpolynomials (and/or the associated `polynomial eigenvalues') of the $D$ operator (present both on the definition for the function $S$ and on for the exponent (Q)). So, the firts job at hand is to determine those. In order to do that we have to choose a degree. Let us, for simplicity sake, start with the most simple possibility. So, the first step is the

\bigskip
{\bf Step 1} - {\it determination of the eigenpolynomials and of the `polynomial eigenvalues' (of degree 1) of the $D$ operator.}

\bigskip
\noindent
When these polynomials are determined and put on equation (\ref{eqadendum5}), the next thing to do is to

\bigskip
{\bf Step 2} - {\it determine the degree for the $Q$ polynomial.} 

\bigskip
\noindent
Please note that the degree for this polynomial is not fully determined by the determination of the degrees for the eigenpolynomials of the $D$ operator, there are still the powers for these polynomials ($m_i$). 

Thus, once the degree for $Q$ is set, we determine all possible values for $m_i$ to accomodate this choice with the choice made on the first step just described. So, the next step is the

\bigskip
{\bf Step 3} - {\it determination of all possible values for $m_i$.}

\bigskip
\noindent
After that step is taken, one can see that, on equation (\ref{eqadendum5}), we still have $P$ to determine. But, from the results present on section (\ref{introM}), from the previous choices we have made, the maximum degree for this polynomial can be determined: we have that the maximum degree for the left hand side of (\ref{eqadendum5}) is $\hbox{\tt d}_{\hbox{\scriptsize {\it P}}} + \max (\hbox{\tt d}_{\hbox{\scriptsize {\it M}}},\hbox{\tt d}_{\hbox{\scriptsize {\it N}}}) - 1$ and the maximum degree for the right hand side of (\ref{eqadendum5}) is $\hbox{\tt d}_{\hbox{\scriptsize {\it Q}}} + \max (\hbox{\tt d}_{\hbox{\scriptsize {\it M}}},\hbox{\tt d}_{\hbox{\scriptsize {\it N}}})~-~1$. Therefore, once the degree of $Q$ is set, the maximum degree for $P$ is $\hbox{\tt d}_{\hbox{\scriptsize {\it Q}}} + \max (\hbox{\tt d}_{\hbox{\scriptsize {\it M}}},\hbox{\tt d}_{\hbox{\scriptsize {\it N}}})$. So we 

\bigskip
{\bf Step 4} - {\it construct a generic polynomial of degree 1 ($P = a_0 + a_1\,x + a_2\,y$).}

\bigskip
Equation (\ref{eqadendum5}) then becomes of the general form:
\begin{equation}
\label{eqcalP}
{\cal P} = 0
\end{equation}
where ${\cal P}$ is a polynomial.  For this to be satisfied, we have to 

\bigskip
{\bf Step 5} - {\it equate the coefficients of the different powers of $(x,y)$ to zero. 
Thus generating a set of linear algebraic equations on the undetermined parameters ($a$'s and $c$'s).}

\bigskip
{\bf Step 6} - {\it Solve the set of linear algebraic equations thus determining the $a$'s and $c$'s.}

\bigskip
If the solution can not be found, we 

\bigskip
{\bf Step 7} - {\it increase the degree of $P$ (up to its maximum degree) and repeat the procedure from {\bf Step 5}.}

\bigskip
If the solution can not be found, we 

\bigskip
{\bf Step 8} - {\it increase the degree of $Q$ and repeat the procedure from {\bf Step 3} onwards.}

\bigskip
Of course, this procedure can go on indefinitely (remember we do not know the bound for the degree of $Q$). However, the degree of the eigenpolynomials (from which we construct $Q$ and $S$) can also be increased (in {\bf Step 1} above it was set to be 1). So, instead of trying all the possibilities for $Q$ (they are infinite!), one can interrupt this arduous loop and experiment starting all over (from {\bf Step 1}) but with an increased value for the degree of the eigenpolynomials. 

By doing that, we are covering all the possibilities and we may hope that we will find a solution within our lifetime. On a brighter tone, most of the examples we have come across are solved with low degree for the eigenpolynomials of the $D$ operator and for $Q$.

\subsection{Examples and Results}

In this section, we are going to present examples of application of our method and discuss its 
effectivness. 

First, in order to illustrate the steps of the method just presented, we are going to start with a simple LFOODE. This example was artificialy manafuctared by us according to two criteria: It is (in a way) simple and it is not solved, as far as we know, by any other method.

\bigskip
\noindent
{\bf Example 1:}

Consider the following LFOODE:
\begin{equation}
{\frac {dy}{dx}}={\frac {\left (x+1\right )y}{x-xy -y^{2}+{x}^{2}}}
\end{equation}

For this equation, up to degree 1 ({\bf Step 1}), we have that the eigenpolynomials (with the associated eigenvalues) are:
\begin{itemize}
\item $v_1 = y,\,\,\,\,\,\lambda_1 = x+1$,  
\item $v_2 = x+y,\,\,\,\,\,\lambda_2 = 1+x-y$.  
\end{itemize}

The next step ({\bf Step 2}) is to choose the degree for the polynomial $Q$. Starting with $\hbox{\tt d}_{\hbox{\scriptsize {\it Q}}} = 1$, since we are in the case where $\hbox{\tt d}_{v_{\!\hbox{\scriptsize {\it q}}}} = 1$ and $\hbox{\tt d}_{v_{\!\hbox{\scriptsize {\it s}}}}= 1$, the only possible values for $m_i$ ({\bf Step 3}) are $\{m_1=1, m_2=0\}$ and $\{m_1=0, m_2=1\}$. For this particular example, the maximum degree for $P$ is 2. So, starting with $\hbox{\tt d}_{\hbox{\scriptsize {\it P}}}=1$ we have $P=a_1+a_2\,x+a_3\,y$ ({\bf Step 4}) and equation (\ref{eqadendum5}) leads to:
\begin{eqnarray}
\label{eqdd0}
{y}^{{\it m1}}\left (x+y\right )^{{\it m2}}\left ({\it n1}\,\left (x+1
\right )+{\it n2}\,\left (1+x-y\right )+3\,x+2-y\right )+\nonumber\\
\left (x-xy-{
y}^{2}+{x}^{2}\right ){\it a2}+\left (x+1\right )y{\it a3}-\nonumber\\
\left ({
\it a1}+{\it a2}\,x+{\it a3}\,y\right )\left ({\it m1}\,\left (x+1
\right )+{\it m2}\,\left (1+x-y\right )\right )
 = 0.
\end{eqnarray}

As we shall see, with the values $m_1=1, m_2=0$ is possible to find a solution. Substituting those into (\ref{eqdd0}) we get:
\begin{eqnarray}
y\left ({\it n1}\,\left (x+1\right )+{\it n2}\,\left (1+x-y\right )+3
\,x+2-y\right )+\left (x-xy-{y}^{2}+{x}^{2}\right ){\it a2}+\nonumber\\
\left (x+1
\right )y{\it a3}-\left ({\it a1}+{\it a2}\,x+{\it a3}\,y\right )
\left (x+1\right )
 = 0.
\end{eqnarray}

In order to solve the above equation, the coefficients for different powers of $(x,y)$ have to be zero. Thus leading to the following set of equations ({\bf Step5}):

\begin{eqnarray}
n1+n2+2=0  \nonumber \\
-n2-a2-1=0  \nonumber \\
-a1=0\nonumber \\
n1+n2+3-a2=0
\end{eqnarray}

Leading to the solution for the coefficients ({\bf Step6}):
\begin{equation}
{a1 = 0, a3 = a3, n1 = 0, n2 = -2, a2 = 1}
\end{equation}

So, the integrating factor for this LFOODE becomes (choosing $a_3=0$):
\begin{equation}
R = \frac{e^{x/y}}{\left (x+y\right )^{2}}
\end{equation}

Then the solution is:
\begin{equation}
C=\frac{y\left (-1+y\right ){e^{x/y}}}{\left (x+y\right )}-{e^{-1}}{\it Ei}(1,-{\frac {x+y}{y}})
\end{equation}

Our method is designed to deal with LFOODEs, so, in order to analyze its effectiveness, we are going to use as our testing ground the LFOODEs found on the book by Kamke \cite{kamke}, a traditional testing arena for ODEs. The LFOODEs are the equations {\bf I.18, I.20, I.27, I.28, 
I.129, I.133, I.146, I.169 and I.235}, as they are numerated in the book by Kamke. These equations and their corresponding Integrating Factors can be found on table \ref{tabela1}. The method deals with all these examples successfuly. In order not to make it too cumbrous for the reader, we are going to present in detail the calculations for just one of those cases, actualy, the most involved one. This choice for the second example was made aiming to show that, even for complex cases, our method is contained and manageable.

\bigskip
\noindent
{\bf Example 2:}

Consider the following LFOODE ({\bf I.169} from the book by Kamke):
\begin{equation}
\left (ax+b\right )^{2}{\frac {dy}{dx}}+\left (ax+b\right )\left (y\right )^{3}+c\left (y\right )^{2}
\end{equation}

For this equation, up to degree 1 ({\bf Step 1}), we have that the eigenpolynomials (with the associated eigenvalues) are:
\begin{itemize}
\item $v_1 = y,\,\,\,\,\,\lambda_1 = -y\,c-b\,y^2-a\,x\,y^2$,  
\item $v_2 = (ax+b)/a,\,\,\,\,\,\lambda_2 = a\,b+a^2\,x$.  
\end{itemize}

The next step ({\bf Step 2}) is to choose the degree for the polynomial $Q$. For this particular example, we will see that $\hbox{\tt d}_{\hbox{\scriptsize {\it Q}}} = 4$ is needed. Since we are in the case where $\hbox{\tt d}_{v_{\!\hbox{\scriptsize {\it q}}}} = 1$ and $\hbox{\tt d}_{v_{\!\hbox{\scriptsize {\it s}}}}= 1$, the possible values for $m_i$ ({\bf Step 3}) are $\{m_1=4, m_2=0\}$, $\{m_1=3, m_2=1\}$, $\{m_1=2, m_2=2\}$, $\{m_1=1, m_2=3\}$ and $\{m_1=0, m_2=4\}$. For this case the maximum degree for $P$ is 8 and the lowest degree to which a solution can be found is 4. So, letting  $P=
a_1\,y^4+a_2\,x^2\,y+a_3\,y^2+a_4\,x\,y^2+a_5\,x\,y+a_6\,x^4+a_7\,x^3+a_8\,x^2+a_9\,y^3+a_{10}\,x^3\,y+a_{11}\,x\,y^3+a_{12}\,x+a_13\,x^2\,y^2+a_{14}\,y+a_{15}$ ({\bf Step 4}), equation (\ref{eqadendum5}) becomes:
\begin{eqnarray}
\label{eqdd}
{y}^{{\it m1}}\left ({\frac {ax+b}{a}}\right )^{{\it m2}}\left ({\it 
n1}\,\left (-yc-b{y}^{2}-ax{y}^{2}\right )+{\it n2}\,\left (ab+{a}^{2}
x\right )-2\,y\left (axy+by+c\right )\right.\nonumber \\
\left.-{y}^{2}\left (ax+b\right )+2\,{a
}^{2}x+2\,ab\right )+\left ({a}^{2}{x}^{2}+2\,axb+{b}^{2}\right )
\left (2\,{\it a2}\,xy+{\it a4}\,{y}^{2}+{\it a5}\,y+\right.\nonumber \\
\left.4\,{\it a6}\,{x}^
{3}+3\,{\it a7}\,{x}^{2}+2\,{\it a8}\,x+3\,{\it a10}\,{x}^{2}y+{\it 
a11}\,{y}^{3}+{\it a12}+2\,{\it a13}\,x{y}^{2}\right )\nonumber \\
-{y}^{2}\left (a
xy+by+c\right )\left (4\,{\it a1}\,{y}^{3}+{\it a2}\,{x}^{2}+2\,{\it 
a3}\,y+2\,{\it a4}\,xy+{\it a5}\,x+3\,{\it a9}\,{y}^{2}\right.\nonumber \\
\left.+{\it a10}\,{x}
^{3}+3\,{\it a11}\,x{y}^{2}+2\,{\it a13}\,{x}^{2}y+{\it a14}\right )-
\left ({\it a1}\,{y}^{4}+{\it a2}\,{x}^{2}y+{\it a3}\,{y}^{2}+{\it a4}
\,x{y}^{2}\right.\nonumber \\
\left.+{\it a5}\,xy+{\it a6}\,{x}^{4}+{\it a7}\,{x}^{3}+{\it a8}\,
{x}^{2}+{\it a9}\,{y}^{3}+{\it a10}\,{x}^{3}y+{\it a11}\,x{y}^{3}+{
\it a12}\,x\right.\nonumber \\
\left.+{\it a13}\,{x}^{2}{y}^{2}+{\it a14}\,y+{\it a15}\right )
\left ({\it m1}\,\left (-yc-b{y}^{2}-ax{y}^{2}\right )+{\it m2}\,
\left (ab+{a}^{2}x\right )\right )
\end{eqnarray}

As we shall see, with the values $m_1=2, m_2=2$ is possible to find a solution. Substituting those into (\ref{eqdd}) we get:
\begin{eqnarray}
\left \{4\,{b}^{2}{\it a6}+6\,ab{\it a7}+2\,{a}^{2}{\it a8}=0,3\,{b}^{
2}{\it a7}+{a}^{2}{\it a12}+4\,ab{\it a8}=0,\right.\nonumber\\
\left.2\,ab{\it a5}+{\it a12}\,c
+{\it n2}\,{a}^{2}+2\,{b}^{2}{\it a2}+2\,{a}^{2}=0,-3\,c{\it a1}-2\,b{
\it a9}=0,\right.\nonumber\\
\left.{b}^{2}{\it a12}=0,-{\it n1}\,c+{\it a15}\,b+{b}^{2}{\it a4}
-2\,c=0,2\,{a}^{2}{\it a2}+{\it a7}\,c+6\,ab{\it a10}=0,\right.\nonumber\\
\left.8\,ab{\it a6}+
3\,{a}^{2}{\it a7}=0,2\,{b}^{2}{\it a8}+2\,ab{\it a12}=0,4\,{a}^{2}{
\it a6}=0,\right.\nonumber\\
\left.-3\,b-{\it n1}\,b+{b}^{2}{\it a11}-c{\it a3}=0,{a}^{2}{\it 
a5}+3\,{b}^{2}{\it a10}+{\it a8}\,c+4\,ab{\it a2}=0,\right.\nonumber\\
\left.b{\it a12}+{\it 
a15}\,a+2\,{b}^{2}{\it a13}+2\,ab{\it a4}=0,-{\it n1}\,a-3\,a-c{\it a4
}+2\,ab{\it a11}=0,\right.\nonumber\\
\left.b{\it a8}+{a}^{2}{\it a4}+a{\it a12}+4\,ab{\it a13}
=0,{\it n2}\,ab+2\,ab+{b}^{2}{\it a5}+{\it a15}\,c=0,\right.\nonumber\\
\left.{a}^{2}{\it a11}-
c{\it a13}=0,-b{\it a3}-2\,c{\it a9}=0,{\it a6}\,a=0,3\,{a}^{2}{\it 
a10}+{\it a6}\,c=0,\right.\nonumber\\
\left.-2\,b{\it a11}-2\,a{\it a9}=0,-3\,a{\it a1}=0,-3\,b
{\it a1}=0,-a{\it a13}=0,-2\,a{\it a11}=0,\right.\nonumber\\
\left.-b{\it a13}-a{\it a4}=0,-2\,
c{\it a11}-a{\it a3}-b{\it a4}=0,\right.\nonumber\\
\left.a{\it a8}+b{\it a7}+2\,{a}^{2}{\it 
a13}=0,a{\it a7}+b{\it a6}=0\right \}
\end{eqnarray}

Leading to the solution for the coefficients:
\begin{eqnarray}
\left \{{\it a6}=0,{\it a11}=0,{\it a10}=0,{\it a1}=0,{\it n2}=-1,{
\it a2}=0,{\it a7}=0,{\it a9}=0,\right.\nonumber\\
\left.{\it a3}=-1/2\,{\frac {{c}^{2}-2\,a{b}
^{2}{\it a13}}{{a}^{3}}},{\it n1}=-3,{\it a14}=-{\frac {cb}{{a}^{2}}},\right.\nonumber\\
\left.{\it a15}=-1/2\,{\frac {{b}^{2}}{a}},{\it a5}=-{\frac {c}{a}},{\it a12
}=-b,{\it a4}=2\,\right.\nonumber\\
\left.{\frac {b{\it a13}}{a}},{\it a13}={\it a13},{\it a8}=
-1/2\,a\right \}
\end{eqnarray}

So, the integrating factor for this LFOODE becomes (choosing $a_3=0$):
\begin{equation}
R = \frac{{a\,e^{-\,{\frac {\left (yc+{a}^{2}x+ab\right )^{2}}{2\,a{y}^{2}\left (ax
+b\right )^{2}}}}}}{{y}^{3}\left (ax+b\right )}
\end{equation}

Then the solution is:
\begin{equation}
C = \int \!-\frac{\left (axy+by+c\right ){e^{\frac{-1}{2\,a}\,\left ({\frac {c}{
ax+b}}+{\frac {a}{y}}\right )^{2}}}}{{y}\left (ax+b\right )}
{dx}
\end{equation}

\section{Conclusion}

In this paper we have present a semi-decision procedure to tackle first order
differential equations, with Liouvillian functions in the solution
(LFOODEs). As in the case of the Prelle-Singer procedure, this method
is based on the knowledge of the integrating factor structure that we
now have shown to be:
\begin{equation} 
\label{ourRconclusion}
R = e^{r_0(x,y)}  \prod_{i=1}^{n} v_i(x,y)^{c_i}.
\end{equation} 
where $r_0=P/Q$, $P$ and $Q$ are polynomials in $(x,y)$, $v_i$ are
irreducible eigenpolynomials of the $D$ operator (where $D \equiv N
\partial_x + M \partial_y$)$, c_i$ are constants, $D[P/Q]$ is a
polynomial in $(x,y)$ and $Q|D[Q]$ $($i.e., $D[Q]/Q$ is a polynomial 
in $(x,y))$.


\end{document}